\documentclass[nologo,11pt,a4paper,hidelinks]{ETHpaper}
\usepackage{graphicx, epsfig, amsmath, amssymb,color,wasysym}
\usepackage[numbers]{natbib}
\usepackage{subcaption}
\usepackage{booktabs}
\usepackage{mathtools}
\usepackage{cancel}
\usepackage[table]{xcolor}
\usepackage{colortbl}
\usepackage{algorithm2e}
\usepackage{dcolumn} 
\definecolor{lgray}{RGB}{210,210,210}
\newcolumntype{d}[1]{D{.}{.}{#1}}

\begin{document}

\newcommand{\mean}[1]{\left\langle #1 \right\rangle} 
\newcommand{\abs}[1]{\left| #1 \right|}

\title{Higher-order models capture changes in controllability of temporal networks}

\titlealternative{Higher-order models capture changes in controllability of temporal networks}

\author{Yan Zhang$^1$, Antonios Garas$^1$, Ingo Scholtes$^{2,3}$ }

\address{
 \mbox{$1$ ETH Z\"urich, Chair of Systems Design, Weinbergstrasse 56/58, CH-8092 Z\"urich, Switzerland}

 \mbox{$2$ University of Wuppertal, Chair of Data Analytics, Gau\ss{}stra\ss{}e 20, D-42119 Wuppertal, Germany}

\mbox{$3$ University of Zurich, Data Analytics Group, Binzm\"uhlestra\ss{}e 14, CH-8050 Z\"urich, Switzerland}}

\reference{(Yan Zhang et al 2020 J. Phys. Complexity, in press)}
\www{\url{http://www.sg.ethz.ch}}

\makeframing
\maketitle

\begin{abstract}

In many complex systems, elements interact via time-varying network topologies.
Recent research shows that temporal correlations in the chronological ordering of interactions crucially influence network properties and dynamical processes.
How these correlations affect our ability to control systems with time-varying interactions remains unclear.
In this work, we use higher-order network models to extend the framework of structural controllability to temporal networks, where the chronological ordering of interactions gives rise to time-respecting paths with non-Markovian characteristics.
We study six empirical data sets and show that non-Markovian characteristics of real systems can both increase or decrease the minimum time needed to control the whole system.
With both empirical data and synthetic models, we further show that spectral properties of generalisations of graph Laplacians to higher-order networks can be used to analytically capture the effect of temporal correlations on controllability. 
Our work highlights that (i) correlations in the chronological ordering of interactions are an important source of complexity that significantly influences the controllability of temporal networks, and (ii)  higher-order network models are a powerful tool to understand the temporal-topological characteristics of empirical systems.

\end{abstract}

\section{Introduction}

Many complex systems have dynamic topologies that can be described by \emph{temporal networks} \cite{Holme2015}. 
Examples range from the metabolic network, where chemical interactions follow a specific ordering, to social networks where the timing of interactions can display a bursty nature.
In such temporal networks, nodes can only influence each other via so-called causal or time-respecting paths \cite {Holme2015}, i.e., sequences of interactions with increasing time stamps that connect nodes.

Previous empirical studies of time series data have shown that the chronological ordering and timing of dynamic interactions often exhibit rich correlations that lead to \emph{non-Markovian} causal paths ~\cite{Lambiotte2018,Scholtes2014a}.
That is, in the sequence of nodes traversed by a causal path, the next node does not only depend on the current node, but also on the previous ones.
The resulting \emph{higher-order dependencies} between nodes imply that, different from paths in a static network, causal paths in a temporal network cannot simply be understood based on the transitive closure of links.
Hence the \emph{causal topology} of such systems, which captures how nodes can indirectly influence each other via causal paths, can be much more complex than what is expected based on the static and aggregated network alone \cite{Rosvall2014,Pfitzner2013,Scholtes2014a}. 
Exploring such higher-order dependencies in temporal networks, recent studies have revealed that they can crucially impact network properties like, e.g. node centralities \cite{Scholtes2016,Xu2016} or community structures \cite{Rosvall2014,Salnikov2016,Peixoto2017}, as well as dynamical processes like, e.g. epidemic spreading \cite{Pfitzner2013,Rosvall2014}, and diffusion \cite{Rosvall2014,Lambiotte2014,Scholtes2014a,Xu2016}.

While these works have helped us to understand how temporal correlations influence a number of network analytic problems, they also raise important questions for controllability, the ability to guide a system towards a desired state by means of suitable control signals.
Analytical and empirical studies of controllability have provided crucial insights into the structure-function relationship of complex systems \cite{Fax2002,Yosef2011,Li2004,Vinayagam2016,Gu2015,Yan2017,Liu2011,Nepusz2012b,zhao2015intrinsic,Yuan2014,Menichetti2016,Sun2013}. 
It allows, for instance, to predict how a certain neuron affects the behaviour of organisms like C. elegans \cite{Yan2017} or to identify disease genes and drug targets in protein interactions \cite{Vinayagam2016}.

The few existing studies on controllability of temporal networks have mainly focused on the comparison of temporal networks with their static counterparts \cite{Li1607}, or addressed the role of the non-Poissonian inter-event time distributions \cite{KOSTIC2014917} on control \cite{Pan2014c,Posfai2014}. 
For example, the authors of \cite{Pan2014c} examined what part of the system can be controlled with a single driver node, finding that both activity patterns and node degrees influence controllability. 
Comparing a temporal network with its static counterpart, \cite{Li1607} showed that the dynamic nature of links can reduce the time needed to achieve full controllability of a system. 
Despite these early explorations, it remains an open question how temporal correlations in the ordering of interactions impact controllability. Additionally, we lack an analytical approach to systematically understand this effect.

In this manuscript we seek to close this research gap.
Our study builds on higher-order models of causal topologies, which have recently been introduced to study the complex interplay between topological and temporal characteristics in real complex systems \cite{Rosvall2014,Pfitzner2013,Scholtes2014a,Lambiotte2018}. 
We explore six empirical time series data sets on social, technical, and biological networks.
We show that order correlations, resulting in time-respecting paths with \emph{non-Markovian} characteristics, can both increase or decrease the minimum time needed to fully control a system. Specifically, we show that, in the same system, temporal correlations can slow down controllability, while they speed up diffusion processes.

Building on higher-order models for causal paths in temporal networks, we further show that the speed-up and slow-down effect observed in empirical data may be captured by the second-smallest eigenvalue of a higher-order generalisation of the graph Laplacian.
To better understand the mechanisms behind the observed effects, we propose a model for temporal networks that allows to tune how temporal correlations affect the causal topology. Our findings provide a possible explanation of why higher-order spectral properties may analytically capture the effect of temporal correlations, thus opening new perspectives to study controllability in temporal networks.

\section{Structural Controllability of temporal networks}

We first define a temporal network as a tuple $G=(V ,E^T)$ with a set of $N$ nodes $V$ as well as a set $E^T \subseteq V \times V \times \mathbb{N}$ of time-stamped links $(i,j;t) \in E^T$, where $t \in \mathbb{N}$ is the discrete time stamp of a (possibly directed) link from node $i$ to $j$.

Such a temporal network can be represented as a series of network snapshots, each snapshot at time $t$ containing only those time-stamped links $(i,j;t)$ occurring at time $t$.
We can represent each snapshot by an adjacency matrix $\mathbf{A(t)} \in \mathbb{R}^{\mathbb{N} \times \mathbb{N}}$, where elements $a_{ij}(t)$ $(i,j=1,...,N)$ capture the presence, and possibly strength, of a weighted link from node $i$ to node $j$ at time $t$.

We further assume that we wish to control a discrete-time linear dynamical process that operates on the nodes of the temporal network.
This assumption of a linear dynamics has been extensively used in recent studies of network control\cite{Li1607,Menichetti2014a,Nepusz2012b,Liu2016}.
Despite its simplicity, and notwithstanding the fact that many real systems exhibit non-linear dynamics, this approach is justified because it is a first-order approximation that can provide analytical insights and reveal important aspects about the interplay between network structure and controllability.
For a vector $\mathbf{X}(t) \in \mathbb{R}^{N}$ describing the state of all $N$ nodes at time $t$, the dynamics of such a linear process on a temporal network can be given as
\begin{equation} \label{eq1}
\mathbf{X}(t+1)=\mathbf{G}(t+1)\mathbf{X}(t)+\mathbf{B}\mathbf{U}(t),
\end{equation}
where matrix $\mathbf{G}(t+1) := \left[\mathbf{A}(t+1)\right]^T+\mathbf{I}$. Matrix $\mathbf{A}(t)$ encodes the time-varying topology of interactions among nodes, and matrix $\mathbf{I}$ captures the assumption of the dynamics that a node keeps its state if it does not interact with others.  
We further assume that the matrix $\mathbf{B} \in \mathbb{R}^{N \times N_d}$ maps a set of \emph{control signals} to a set of $N_d$ \emph{driver nodes}, which can be directly controlled.
In particular, we assume that the matrix $\mathbf{B}$ maps a time-varying vector $\mathbf{U}(t)$ of $N_d$ control signals $u_{j}(t)$ $(j=1,2,..,N_d)$ to $N_d$ driver nodes, i.e. we assume that $b_{ij} \neq 0$ iff input signal $u_j$ is assigned to driver node $i$.

Having defined the temporal network and the dynamical process, we now formalise the notion of \emph{controllability} of this process in a temporal network.
We can actually view the evolution of such a process as a trajectory in an $N$-dimensional state space, where $N$ is the number of nodes.
Our ability to control the state of $k$ of these $N$ nodes corresponds to the ability to design suitable control signals such that we are able to guide the trajectory to an arbitrary point in a $k$-dimensional subspace.
Following the algebraic approach introduced by Kalman~\cite{Kalman1963}, the size of this controllable subspace (i.e. the number of controllable nodes) of a linear dynamical system for a given set of driver nodes can be assessed by calculating the rank of a so-called \emph{controllability matrix}.
For our scenario of a dynamical system in a temporal network, the common definition of this matrix naturally leads to the following \emph{temporal controllability matrix}~\cite{Liu2013,Pan2014c,Posfai2014}
\begin{equation} \label{controllabilityMatrix}
\mathbf{C}_{t}=[\mathbf{G}_{t}\mathbf{G}_{t-1}...\mathbf{G}_{1}\mathbf{B}, \mathbf{G}_{t}\mathbf{G}_{t-1}...\mathbf{G}_{2}\mathbf{B},\ldots, \mathbf{G}_t\mathbf{B},\mathbf{B}] \in \mathbb{R}^{N \times t N_d},
\end{equation}
where $[\mathbf{A},\mathbf{B}]$ denotes the concatenation of two matrices $\mathbf{A}$ and $\mathbf{B}$ and the products $\mathbf{G}_{t} \cdot \ldots \cdot \mathbf{G}_{1}$ take the role of the matrix power $\mathbf{A}^t$ in the common definition of the controllability matrix for static networks~\cite{Kalman1963}.
It has been shown that $N_b:=\text{rank}(\mathbf{C}_t)\leq N$ gives the size of the controllable (sub)system at time $t$ for a given mapping of control signals to driver nodes captured in $\mathbf{B}$.
Moreover, using the so-called Kalman ranking condition, we call the system controllable if the temporal controllability matrix has full rank, i.e. all nodes in the temporal network can be controlled~\cite{Liu2013}.

In general, the study of controllability based on the rank of the controllability matrix introduced in Eq.\ref{controllabilityMatrix} allows to incorporate \emph{weighted} links, where weights capture the \emph{strengths} of interactions between nodes.
However, in many real world situations -- including the data sets studied in this manuscript -- the weights of links, or strengths of interactions, are unknown. 
In~\cite{Lin1974} this problem has been addressed based on the framework of \emph{structural controllability}.
The key idea is to treat both the adjacency matrix $\mathbf{A}$ and the ``mapping'' matrix $\mathbf{B}$ as \emph{structural matrices} whose non-zero elements are treated as free parameters.
We then call a system ``structurally controllable'' iff we can tune the free parameters in the structural matrices $\mathbf{A}$ and $\mathbf{B}$ such that the rank $N_b$ of $\mathbf{C}$ equals $N$.
In a recent work, \cite{Liu2011} developed analytical tools to address the controllablity of static networks, which allows to identify the minimum number of drivers based on the network structure of the system.

To apply the concept of structural controllability to temporal networks, \cite{Posfai2014} proposed the \emph{independent path theorem}. 
This theorem states that a set $C$ of nodes in the system is structurally controllable at time $T$, if there exist $|C|$ \emph{independent} paths starting from any input signal to every node in $C$ at time $T$.
Under this condition, it has been shown that the size $N_b$ of the maximally controllable subsystem equals the maximum number of independent paths~\cite{Posfai2014}.
In \cite{Posfai2014}, one important assumption is that \emph{all} link weights are free parameters. 
However, this assumption does not hold in our case because the existence of the identity matrix in $\mathbf{G}$, which corresponds to a set of links of fixed weights. 
To account for this difference, we further extend the \emph{independent path theorem} as we elaborate in the appendix. 
Based on our extension, identifying the size of the controllable subsystem can be similarly mapped into a maximum flow problem that can be solved efficiently in polynomial time~\cite{Goldberg1988}.

\begin{figure}
	\centering
	\includegraphics[width=0.2\columnwidth]{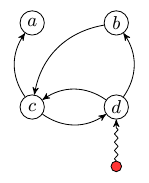}
	\includegraphics[width=0.33\columnwidth]{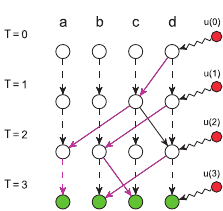}
	%
	\includegraphics[width=0.33\columnwidth]{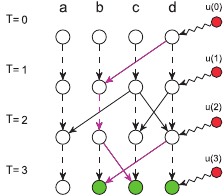}
	\put(-310,-20){\bf{(a)}}
	\put(-220,-20){\bf{(b)}}
	\put(-80,-20){\bf{(c)}}

	\caption{Controlling a simple time varying system of four nodes with only one input signal.
		(a) Aggregated representation of the system.
		(b-c) Two temporal networks with the same aggregate topology as in panel (a), but different ordering of links.
		Red nodes denote control signals. Solid links denote interactions among nodes, and dashed links represent the identity matrix in $\mathbf{G}$. In each of the two temporal networks, we highlight one maximal set of independent time respecting paths in purple, and highlight the corresponding controlled subsystem in  green.
		In panel (b), every node is controllable at time $t=3$, so controllability of the whole system is achieved after three time steps.
		As comparison, only three nodes are controllable at time $t=3$ in panel (c).
		The difference in the size of the controllable (sub)system shows that the chronological ordering of links affects the controllability of time varying systems.}
	\label{TUN}
\end{figure}

As a toy example, Figure.~\ref{TUN} illustrates the notion of controllability in temporal networks, as well as its dependence on independent paths.
Figure.~\ref{TUN} (b) and (c) contain time-unfolded representations of two different temporal networks which are consistent with the same time-aggregated topology shown in Figure.~\ref{TUN} (a).
In this example, we are interested in the controllability of the four nodes at time $T=3$, considering a single driver node $d$ receiving a time-varying input signal $u_t$.
Links that belong to an independent causal path from this driver node to one of the temporal copies at $T=3$ are highlighted in purple.
In Figure.~\ref{TUN} (b), all four temporal copies at time $T=3$ are endpoints of independent causal paths and thus the whole system can be controlled at $T=3$.
In contrast, in Figure.~\ref{TUN} (c), only three of the four nodes can be controlled, since node $a$ is not an endpoint of an independent causal path originating at a driver node.
Since the temporal networks in Figure.~\ref{TUN} (b) and (c) have the same time-aggregated topology, 
this simple example further highlights that the ordering of links influences the controllability of dynamical processes in temporal networks.


\section{Controllability of Empirical Temporal Networks}

We now investigate how the chronological ordering of links affects controllability  in six empirical temporal networks:
(AN) captures 1,911 time-stamped and directed antenna-antenna interactions between 89 individuals in an ant colony~\cite{Blonder2011}.
(RM) contains 26,260 recorded proximity relations between 64 students and academic staff in a university campus~\cite{Eagle2006a}.
(EM) captures 11,000 e-mails exchanged between 167 employees in a manufacturing company for one month~\cite{Editors2010}.
(HO) contains more than 15,000 time-stamped contacts recorded by proximity sensing badges among 46 healthcare workers and 29 patients in a hospital for 48 hours~\cite{Vanhems2013}.
(FL) includes 230,000 multi-segment flights among 116 US airports in the fourth quarter of 2001~\cite{bts}.
(LT) contains itineraries of passengers using the 309 London Underground stations, extracted from the Rolling Origin and Destination Survey database that covers four million passenger flows for one week~\cite{londontube}.

To explore to what extent the time ordering influences controllability, we compare each of the empirical data sets with a randomised temporal network.
This randomised version is identical in terms of topology, frequency, and number of interactions, except for the fact that we have randomly shuffled the time stamps of all interactions.
We obtain a ``null model'' in which all correlations in the ordering of interactions have been removed, while the resulting time-aggregated network is the same as in the empirical data.
To quantify the effect of link ordering on controllability, we calculate the relative size of the subsystem $n_b(T)=N_b(T)/N$ that can be controlled at any time $T$ both for the empirical and the shuffled data.
To facilitate the comparison, we use the same set of randomly sampled driver nodes for each time step, i.e. we choose a set of driver nodes once and then calculate the size of the controllable sub system for (i) the empirical interaction sequence, and (ii) one shuffled sequence for all times $T$.
We repeat this procedure $100$ times for a fixed fraction of randomly chosen driver nodes of 10\%.
As we will show later with a toy model, our results do not qualitatively depend on the choice of this random fraction.

\begin{figure}[t]
	\centering
	\includegraphics[width=0.99\columnwidth]{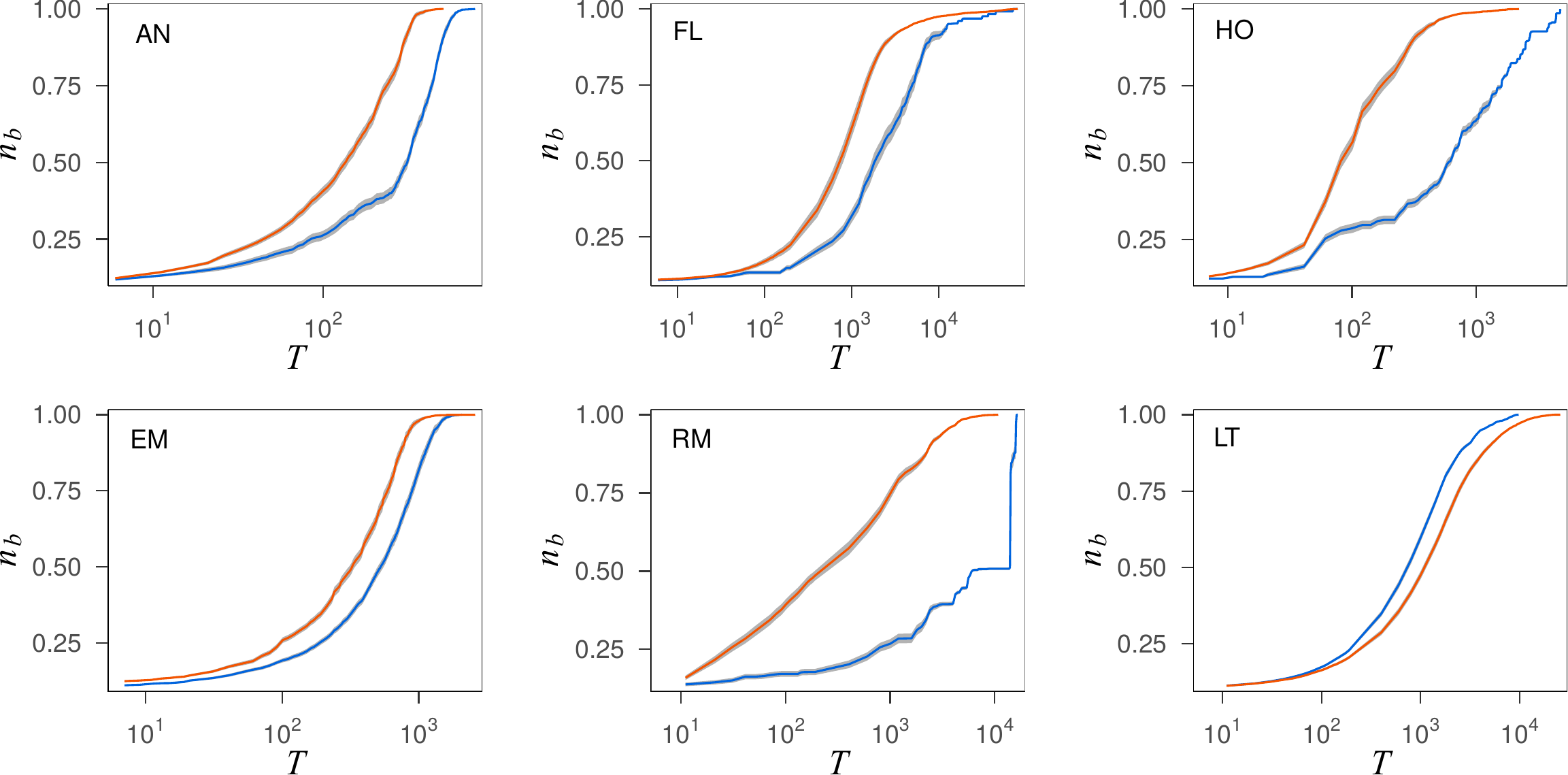}
	\caption{Relative size of the controllable system $n_b(T)$ at time $T$, where a random sample of $10 \%$ of nodes are used as driver nodes.
		Blue dashed lines correspond to the original interaction sequences, the orange continuous lines correspond to shuffled interaction sequences.
		The shaded areas indicate the 95\% confidence intervals of the mean for $100$ realizations.}
	\label{res:controllableSize}
\end{figure}

Figure.~\ref{res:controllableSize} shows the relative size $n_b(T)$ of the subsystem controllable at time $T$ for each of the six empirical data sets along with the corresponding shuffled versions.
The hull curve shows the 95\% confidence interval of the mean value for $100$ simulations of the procedure above.
The fact that the confidence intervals are barely visible for all six cases in Figure.~\ref{res:controllableSize} confirms that the choice of the precise set of driver nodes does not strongly influence our results.
Whenever the size $n_b(T)$ of the controllable subsystem for the empirical data set is smaller than that for the randomized version at a given $T$, the chronological ordering of links negatively affects the size of the controllable subsystem, and vice versa.
Our results show that for all six data sets the ordering of interactions significantly influences controllability.
More precisely, for five of the six data sets the size of the controllable subsystem grows slower due to correlations in the chronological ordering, while for one case (LT) these correlations speed up controllability. Here we highlight that this observation is different from \cite{Scholtes2014a}, which uses the same six datasets but focuses on a different dynamics. We will elaborate this point later in Section. \ref{sec:higherorder}.

\begin{figure}[t]
	\centering
	\includegraphics[width=0.99\columnwidth]{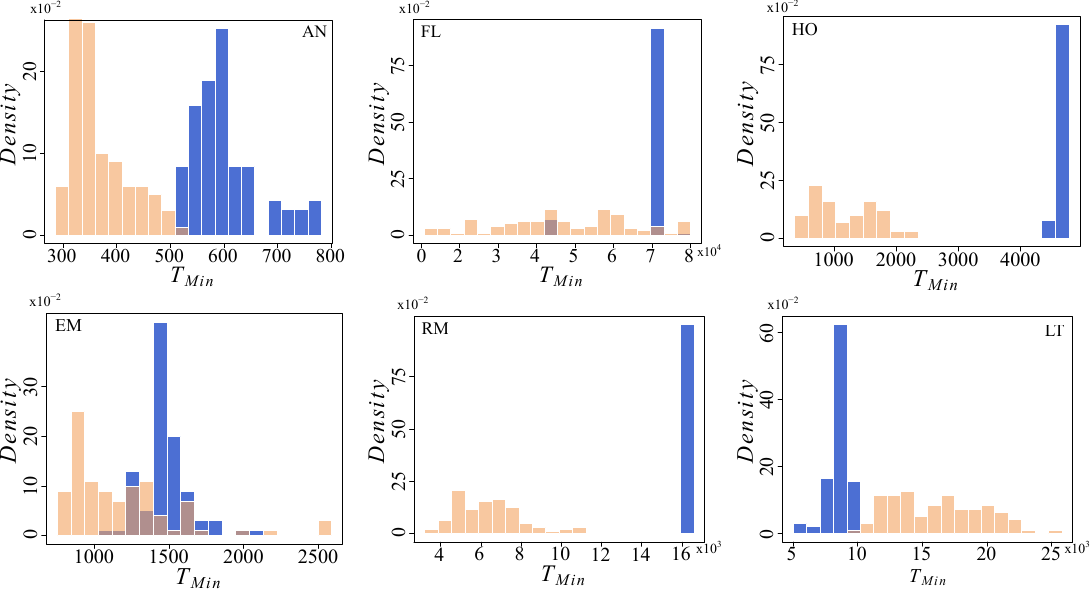}
	\caption{Distribution of the minimum time $T_{Min}$ required to achieve controllability of the whole system (i.e. $n_b(T_{Min})=1$).
		The blue bars refer to the original interaction sequences, and the orange bars to the shuffled interaction sequences.}
	\label{res:Time}
\end{figure}

Besides the size of the controllable subsystem at a given time $T$, we can investigate the minimum time $T_{Min}$ required to control the full system, i.e. $T_{Min} := \arg \min_T n_b(T)=1$.
Figure.~\ref{res:Time} compares the distribution of $T_{Min}$ for each of the data sets to its shuffled counterparts.
As before, for five of the six data sets the peak of the $T_{Min}$ distribution in the shuffled sequence is shifted to the left compared to the empirical data, thus indicating that the ordering of interactions slows down controllability.
On the other hand, for (LT) the peak of the distribution for the shuffled sequence is shifted to the right, thus indicating a speed-up of controllability.
These results show that the ordering of interactions in temporal networks can both speed up and slow down controllability.

\section{Higher-Order Analysis of Controllability}
\label{sec:higherorder}

Our finding that the chronological ordering of links alone can either make it harder or simpler to control a  system raises the question whether this effect can be explained or even predicted. This is a difficult question, because the condition for achieving controllability is much more rigid than previously studied dynamics. 
Therefore, to answer this question, we exploratorily utilize the higher-order modelling approach introduced in~\cite{Scholtes2014a}. 
The idea behind is to construct static, time-aggregated representations of temporal networks that encode information on both the topology and the chronological ordering of time-stamped links.
Here we limit our study to second-order representations, which are the simplest possible higher-order generalisation of (first-order) time-aggregated networks. 
Despite this simplicity, second-order network models capture how the ordering of links affects a system's \emph{causal topology} and thus dynamical processes~\cite{Scholtes2014a}.

Following ~\cite{Scholtes2014a},  a first-order time-aggregated network is constructed by aggregating all time-stamped links that occur in a temporal network, i.e. we simply discard all information on the timing and ordering of links.
Intuitively, the weight of a link in this static representation can be defined as the number of times it appears in the temporal network.
Building on the observation that the weights of links in such \emph{first-order} time-aggregated representations count the frequency of links -- and thus causal paths of length one -- we can generalise this approach to second-order network models that account for higher-order dependencies.
A \emph{second-order} network model can be constructed following a line graph construction:
First of all, each link $(a,b)$ in the first-order network defines a node $a-b$ in the second-order network.
Two second-order nodes $a-b$ and $b-c$ are connected by a directed second-order link $(a-b, b-c)$, if the corresponding causal path $a \rightarrow b \rightarrow c$ of length two exists. 
We additionally assume a limited waiting time $\delta$ for causal paths, i.e. we assume that a path $(a-b,b-c)$ exists if there are two time-stamped link $(a,b;t_1)$  and $(b,c;t_2)$ such that $0<t_2-t_1<\delta$. 
The parameter $\delta$ can be thought as capturing the time scale of the dynamical process that we study.
Moreover, we define the weight of the link $(a-b, b-c)$ to capture the frequency of the causal path $a \rightarrow b \rightarrow c$  in the temporal network.


As argued in \cite{Scholtes2014a,Scholtes2016,Lambiotte2018}, a major benefit of such higher-order models is that they better capture the \emph{causal topology} of temporal networks, which helps to improve the modelling of dynamical processes as well as network analytic methods.
Previous studies have shown that the effect of the ordering of links on the causal topology can be understood analytically based on the spectral properties of higher-order generalisations of graph Laplacians~\cite{Scholtes2014a}.
Because controllability is a propagation of independent control signals, we follow a similar approach by using spectral properties of higher-order matrix representations to quantitatively assess how ``connected'' the causal topology of a temporal network is.
Intuitively, the ``better connected'' a temporal network is, the faster the propagation of control signals and the faster we are able to achieve control of the the full system.
This level of connectivity is captured by the algebraic connectivity of the network topology, which is defined as the second-smallest eigenvalue, $\lambda_2$, of its Laplacian matrix.
Additionally, the time required to achive full controllability is the first time when every node receives at least one control signal, which is similiar to the first hitting time of a random walk process, which can also be captured by  $\lambda_2$ \cite{Lin2013}.  
Despite that achieving network controllability is more rigid than a diffusion or random walk process,  based on the above reasoning, we first hypothesise that the slow-down or speed-up of controllability observed in the empirical data sets, can be explained by changes in the algebraic connectivity of second-order networks.

\begin{figure}[t]
	\centering
	\includegraphics[width=1\columnwidth]{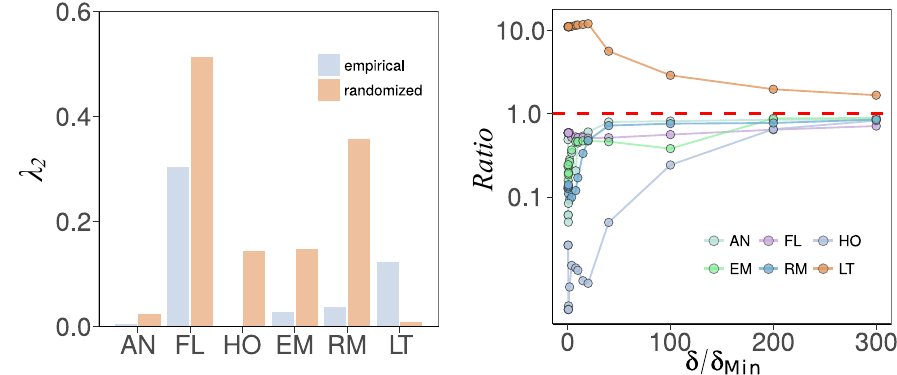}
		\put(-310,-10){\bf{(a)}}
		\put(-160,-10){\bf{(b)}}

	\caption{(a) Algebraic connectivity $\lambda_2$ of the second-order network of the empirical interaction sequences (blue) and a shuffled temporal sequence without order correlations (orange). (b) Ratio of  $\lambda_2$ of the empirical sequences over that of shuffled sequences for different $\delta$.}
	\label{res:algebraic}
\end{figure}

Figure.~\ref{res:algebraic}(a) compares the algebraic connectivity $\lambda_2$ of the second-order network for each empirical data set to the algebraic connectivity of its shuffled counterpart.
We notice that for the five cases where we observed a slow-down in controllability, $\lambda_2$ for the empirical network is smaller than for the shuffled version.
For the (LT) data set, which is the only case in which we observed a speed-up, $\lambda_2$ for the empirical network is larger than for the shuffled version. 

The results above, have been obtained for the smallest time difference $\delta_{\text{Min}}$ used in the definition of causal paths such that the system is still temporally connected, i.e. all nodes can influence each other via causal paths.
To illustrate that our results does not depend on this specific choice of the parameter $\delta$, we compute the ratio of $\lambda_2$ of the second-order time-aggregated network for the empirical data with that of the shuffled counterpart for different values of $\delta$.
Figure.~\ref{res:algebraic}(b) shows that, despite fluctuations for $\delta<\delta_{Min}$, for each dataset, the corresponding curves always stay below or above one, which indicates whether it is easier or harder to achieve full control.
In consequence, we hypothesise that comparison of the algebraic connectivity of a second-order representation of a temporal network with its shuffled counterpart can qualitatively capture how the ordering of links, and thus the resulting changes in the causal topology of a system, affect controllability in the six studied data sets.

\section{Validation in a Synthetic Model}
One could still argue that the above findings that the algebraic connectivity of the second-order network captures the speed-up and slow-down effect are a matter of coincidence in the data. To further support our findings, we introduce a synthetic toy model.
This model is constructed in the spirit of the Watts-Strogatz model~\cite{Watts1998}, in which the algebraic connectivity can be changed by mitigating or enforcing specific paths in the second-order network. 
Concretely, this model generates temporal sequences based on a two-dimension lattice, in which long-range edges are introduced by rewiring. 
A free parameter $\alpha \in [-1,1]$ alters the algebraic connectivity of the second-order network by tuning whether the long-range edges are enforced ($\alpha>0$) or mitigated ($\alpha <0$).
A long-range edge is enforced if it is more likely to appear than expected in the temporal paths that connect two distant nodes in the lattice. 
The Markovian case corresponds to $\alpha=0$, in which the long-range edges are neither enforced nor mitigated. 
For each $\alpha$, we generate one set of temporal sequences based on the second-order network, and one set of the corresponding shuffled sequences. 
A detailed description of the model can be found in the appendix.

\begin{figure}[t]
	\centering
	\includegraphics[width=0.95\columnwidth]{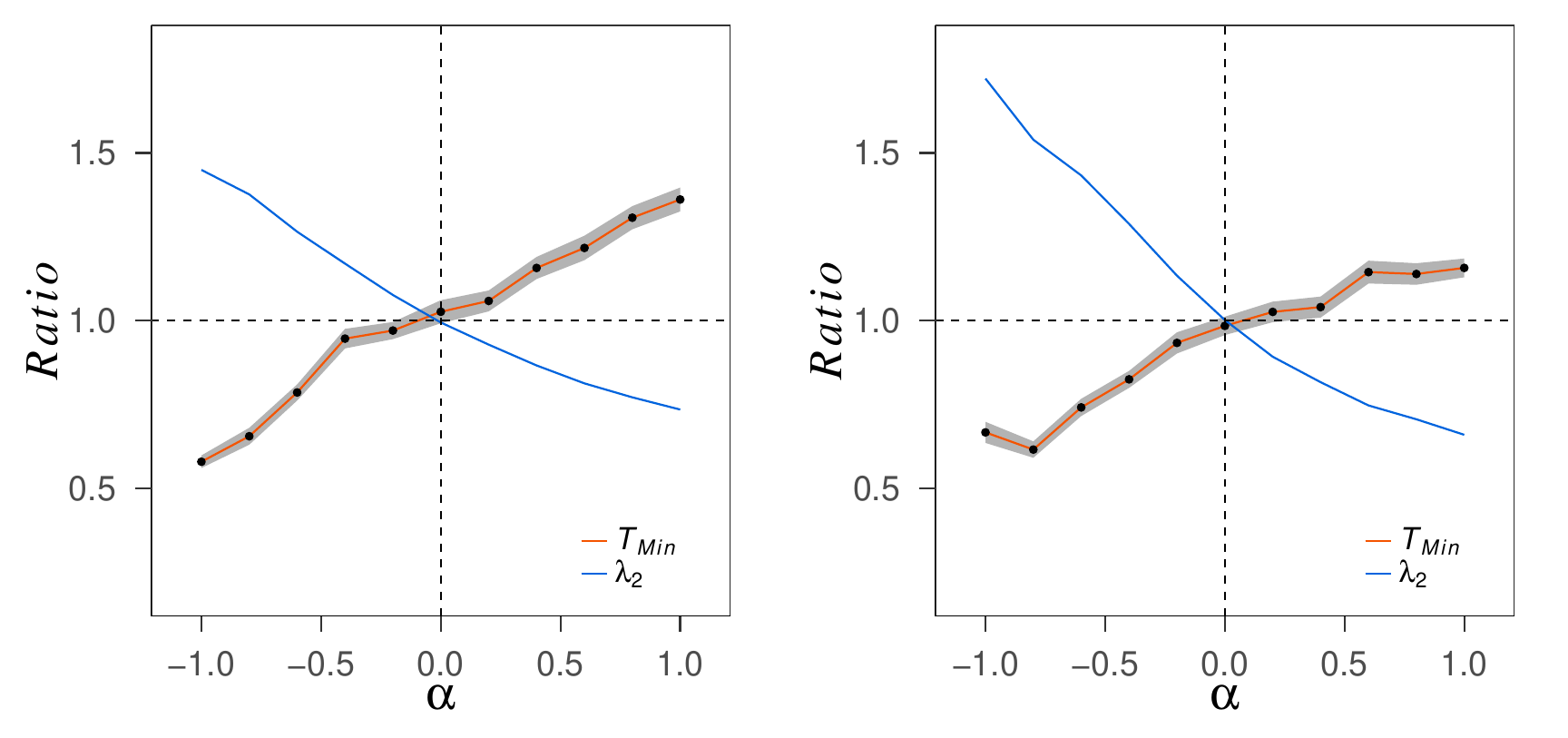}
		\put(-310,-10){\bf{(a)}}
		\put(-160,-10){\bf{(b)}}

	\caption{We study a synthetic model constructed from a lattice with rewired edges. In this model, we consider a lattice of $N=100(400)$ nodes, and each node is connected to its nearest 4 neighbors. Edges in this lattice are rewired with a probability $p=0.1$. Varying the free parameter $\alpha$, we have different second-order networks. For each, we generate 100 interaction sequences with $50000$ events, together with the corresponding shuffled counterpart. Based on these interaction sequences, we look into the ratio of the algebraic connectivity $\lambda_{2}$(blue) and the minimum time to achieve full controllability $T_{Min}$(orange) under different $\alpha$s, with (a) 20\% and (b) 30\% nodes as drives. The gray ribbon shows the 95\% confidence interval, and the dashed lines show the boundary of regions for the speed-up and slow-down effect.} 
	\label{si:toymodel}
\end{figure}

Figure.~\ref{si:toymodel} shows the effect of $\alpha$ on the ratio of the algebraic connectivity $\lambda_{2}$ and the minimum time to achieve controllability $T_{Min}$ between the above two sets of temporal sequences. 
When $\alpha$ is non-zero, the ratios of $\lambda$ are away from the horizontal dashed line, indicating the presence of order correlations in temporal sequences.
When $\alpha$ is smaller than 0, temporal paths with long-range edges that connect two distant nodes are mitigated and the time to achieve controllability is increased compared with the shuffled version. 
In comparison, with $\alpha$ larger than 0, long-range edges are favored in temporal paths that connect distant nodes, and it is faster to achieve controllability of the whole system. 
This way, using simulation results that have been obtained based on a synthetic model, we show that the speed-up effect is not only limited to the (LT) dataset, but it can be observed if the long-range edges are enforced in the system.
Also, we show that the algebraic connectivity of the second-order network captures the speed-up and slow-down effect. This also implies, despite that achieving controllability requires a more rigid condition than a diffusion process, the connectivity of the causal topology is one of the most important factors that influence the slow-down and speed-up effect.

As a final remark, we contrast our findings with the results presented in \cite{Scholtes2014a}, which used the spectral analysis of causal topologies to study the effect of order correlations on the speed of a diffusion process.
Interestingly, our results highlight that the effect of link ordering on diffusion dynamics and control can be different with the same data set.
In particular, our study reveals that the ordering of links in (FL) slows down controllability, while the analysis in \cite{Scholtes2014a} has revealed a speed up of diffusion dynamics.
These opposite effects can be intuitively understood by considering that the speed of diffusion is related to the convergence time of a random walker, while the emergence of controllability is related to the time at which nodes are first reached by a control signal.
From an algebraic point of view, this intuition is captured by the fact that the \emph{speed up} of diffusion in (LT) can be analytically explained based on the spectral gap of a transition matrix~\cite{Scholtes2014a}, which captures the convergence time of a random walker.
In contrast, in our work we have shown that the \emph{slow down} of controllability can be captured based on the algebraic connectivity, which has been shown to predict the first hitting time of a random walker \cite{Lin2013}.

\section{Conclusion}

In summary, we have investigated how the temporal correlations in the chronological ordering of interactions impact controllability, and  developed an analytical approach to systematically understand this effect.
Applying structural controllability theory to six empirical data sets, we showed that temporal correlations can both increase or decrease the minimum time needed to make a system fully controllable. 
Counter-intuitively, we found that even in the same system, the ordering of interactions can have opposite effects on different dynamical processes. 
Furthermore, constructing higher-order network models for causal topology in temporal networks, we showed that the speed-up and slow-down effect observed in empirical data may be captured by the algebraic connectivity of the second-order network. This was confirmed with a synthetic stochastic model that generates interaction sequences with temporal correlations.  

While our analysis of this synthetic model confirms the intuition that the slow-down and speed-up effects observed in empirical temporal networks are associated with the algebraic connectivity of a higher-order model of causal paths, we emphasize that our study does not rigorously establish a direct causal relation between the causal topology in temporal networks and our ability to efficiently control a (linear) dynamical process . In particular, our results do not allow us to predict the magnitude of the slow-down or speed-up effect based on the algebraic connectivity, which is an interesting problem left for future work. We should also be careful in claiming that our findings generalize to other empirical data or synthetic models and thus call for further experiments in data sets with other (temporal) characteristics. To the best of our knowledge, our work is the first suggesting the use of higher-order network models to control complex systems with dynamic interaction topologies. It thus opens new perspectives for the analytical study of the controllability of complex systems and paves the way for future works in different directions:
One direction is to explore how chronological ordering  of interactions influences other aspects of control, such as the minimum set of driver and control energy, if we are able to formulate the actual dynamics \cite{Yan2015} with the link weights precisely measured. 
Another possibility is to look into how we can use model selection techniques to identify the optimal order for the higher-order representation of interaction sequences. In our work, we limit our analysis to second-order models, which are the simplest models to capture chronological ordering  of interaction sequences. Despite this simplicity, it can already highlight how higher-order models can be used to capture the speed-up and slow-down effect. 
It also deserves further effort to find empirical evidence to support our observations on the controllability of temporal networks. 
Answers to these questions will help to understand the fundamental principles behind controlling temporal networks.

\section{Acknowledgements.}

All authors designed the research. 
Y.Z. analysed the empirical data and did the analytical and numerical calculations. 
Y.Z., A.G. and I.S. wrote the manuscript.
We gratefully acknowledge discussions with Frank Schweitzer (Chair of Systems Design, ETH Z\"urich) and Yuan Lin (School of Computer Science, Fudan University) on an early draft of this work.
We are further grateful for helpful comments by Yang-Yu Liu (Harvard Medical School).
A.G. acknowledges support from the EU-FET project MULTIPLEX 317532.
I.S. acknowledges support from the Swiss State Secretariat for Education, Research and Innovation (SERI), Grant No. C14.0036, from the MTEC Foundation project ``The Influence of Interaction Patterns on Success in Socio-Technical Systems'', and from the Swiss National Science Foundation (SNSF), grant 176938.
We declare no competing financial interests.

\section{Data availability statement}
The data that support the findings of this study are available upon request from the authors.

\section{Appendix}

\subsection{Processing datasets}
We study controlllability of temporal networks with six datasets, which have also been used in \cite{Scholtes2014a}. 
To show that the choice of $\delta$ in constructing second-order network has no impact on the result, we use the raw data instead of granulated temporal links. 
We process (FL) and (LT) data sets following the same procedures as indicated in \cite{Scholtes2014a}. For the rest four data sets, we choose the smallest $\delta$ so that most of the nodes in the second-order network can mutually reach each other through time-respecting paths, and we only use temporal links among nodes in the strongly connected component. This way, we remove nodes only appear few number of times in the data set that can hardly reach others or be reached by temporal paths. For the (AN) data set, we set $\delta_{min}=7$ second, so that we have a strongly connected component with 68 nodes. For the (RM) data set, we have $\delta_{min}=300$ seconds, and the resulting dataset contains 83 individuals. For the (HO) dataset, we choose $\delta_{min}=60s$, and we have interactions among 63 individuals. For the (EM) data set, we set $\delta_{min}=30$ minutes, this results a subset of 94 employees.  Note that we also run our analysis with the granulated temporal links as those exactly used in \cite{Scholtes2014a}, which does not change our main results.

\subsection{Constructing a lattice-based synthetic model} 

Our synthetic model starts from a two dimensional-lattice of $N$ nodes, where long-range edges are introduced by rewiring each edge with a probability $p$.
Based on this rewired lattice as the first-order network, we construct the second-order network. Each node in the second-order network corresponds to one edge in the first order. 
For any two nodes $i-j$ and $j-k$ in the second order network, we assume that the weight of the edge connecting these two nodes $w_{i-j,j-k}$ depends the distance between $i$, $j$ and $k$. Particularly, we define $w_{i-j,j-k}=(\frac{(d_{i,k}+1)}{(d_{i,j}+1)(d_{j,k}+1)})^\alpha$, where $d_{i,j}$ denotes the euclidean distance between node $i$ and $j$ on the lattice, and $\alpha \in [-1,1]$ is a free parameter that enforces or mitigates long-range edges. A long-range edge $i-j$ is enforced, if the temporal path $i \rightarrow j \rightarrow k$ is more likely to appear in all the temporal sequences that connect two nodes $i$ and $k$, compared with the temporal sequences expected from the first-order lattice.  
Finally, to generate temporal sequences, we simulate a random walk process on the weighted second-order network: starting from a randomly chosen node in the second-order network, the next node to visit in each step is chosen with probabilities proportional to the weights of edges.


\subsection{Structural controllability of temporal networks}
This work applies structural controllability to temporal networks. 
Note that structural controllability is only one of several structure-based approaches to study network control~\cite{Nacher2016,Mochizuki2013,Zanudo2016,LX2019}.
With the convenience of being mathematically simple, structural controllability theory has provided insights into real control questions, with its predictions validated recently by experimental results~\cite{Vinayagam2016,Gu2015,Yan2017}.
Additionally, the assumption of a linear dynamical process allows structural controllability to be extended to temporal networks, mapping it to a graphical problem that can be solved efficiently~\cite{Posfai2014}.
Since this is not the case for other structure-based approaches, we focus on the structural controllability framework.

The generalisation of this framework to temporal networks involves two steps:
In the first step, we project the time-varying network topology into a so-called \emph{time-unfolded network}, a static directed acyclic graph in which time is ``unfolded'' into an additional topological dimension~\cite{Pfitzner2013,Posfai2014}.
In the second step, we study structural controllability in the temporal network by casting it into a graph-theoretic problem on this static representation.

As a first step, we generate a time-unfolded representation of a temporal network as follows:
For a given set of nodes $V$ and time stamps $\left[1, \ldots, T \right]$ we create ``temporal copies'' $v_t$ for all nodes $v \in V$ and time stamps $t \in \left[1, \ldots, T \right]$ as illustrated in Figure.~\ref{TUN}.
Moreover, for each time-stamped link $(v,w;t)$ we generate a directed \emph{interaction link} $(v_t, w_{t+1})$ connecting the temporal copy of $v$ at time $t$ with the copy of $w$ at time $t+1$.
We obtain a directed acyclic graph in which time moves from top to bottom.
This allows us to study \emph{causal} or \emph{time-respecting paths}~\cite{Pan2011a} as \emph{static} paths in a directed acyclic graph.
In addition, we introduce so-called \emph{state persistence links}, which for each node $v$ connect consecutive temporal copies $v_t$ and $v_{t+1}$ by a directed link $(v_t, v_{t+1})$.
As we explain in more detail later, state persistence links (dashed links in Figure.\ref{TUN}) ensure that the state of a node at time $t$ is \emph{transferred} to the next time step $t+1$.
Without these additional links, for such data sets the state of a temporal copy $v_t$ would be zero whenever there are no links from a previous time step.
We finally add control signals $u_{k}$ connected to all driver nodes $k=1,2,...,N_d$ at every time step $t$.
This time-unfolded projection allows us to address structural controllability of temporal networks through static network analysis.

In order to address structural controllability of the above time-unfolded network, one initial idea to to apply the 
\emph{independent path theorem}~\cite{Posfai2014}, 
which states that a set $C$ of nodes in the system is structurally controllable at time $T$, if there exist $|C|$ \emph{independent} paths starting from any input signal to every node in $C$ at time $T$. 
However, we can not apply this theorem to our setting directly. The reason is as folows:
In the original proof of the theorem \cite{Posfai2014}, it assumes that \emph{all} link weights are free parameters. 
With this assumption, applying the notion of \emph{stem-cycle disjoint subgraphs} as used in~\cite{Liu2011}, then every node $\mathbf{\hat{C}}$ in a stem-cycle disjoint subgraph of the time unfolded network is structurally controllable. Because  $\mathbf{C}$ is a subset of $\mathbf{\hat{C}}$, then every node in $\mathbf{C}$ is controllable. 
However, in the case with state persistence links of fixed weight, not every node in $\mathbf{\hat{C}}$ is controllable. Therefore, we cannot directly conclude every node in $\mathbf{C}$ is controllable.

To overcome this problem, we adapt the structural controllability framework in such a way that it accounts for the special semantics of \emph{state persistence links} and \emph{temporal copies} in time-unfolded networks. 
We first note that, due to the directedness of time, time-unfolded networks are necessarily acyclic, i.e. their stem-cycle disjoint subgraphs are sets of disjoint stems with no cycles. 
Since weights of state persistence links cannot be treated as free parameters, we cannot directly conclude that every node in the stem-cycle disjoint subgraph of a time-unfolded network is structurally controllable.
However, to be able to control the state of all nodes at time $T$, it is not required to control all \emph{temporal copies} of these nodes in all time steps $t<T$. 
This means that we do not require all nodes in the stem-cycle disjoint subgraph to be structurally controllable.
At the same time, if we are able to control a temporal copy $v_t$ on a stem, we can additionally control at least one of the downstream nodes on this stem for time $t'> t$. 
Answering the question whether the nodes $v_T$in a set $C$ are controllable, thus translates to the problem of finding a set of disjoint stems such that each node $v_T$ in $C$ is the endpoint of a stem.
As such, we are only interested in the question whether we can find a set of disjoint stems such that each node $v_T$ in $C$ is the end node of a stem.
Notably, this set of \emph{disjoint} stems corresponds to a set of \emph{independent} causal paths between driver nodes and the nodes in $C$, i.e. a set of causal paths whose nodes are not overlapping. 
This way, we extend the \emph{independent path theorem} so that it can be applied to the case with state persistence links.  

As a brief summary, here we contrast our work with  \cite{Posfai2014}: The assumptions in the dynamics are different as indicated in the form, in our work, we assume that a node keeps its state if it does not interact with others. However, in \cite{Posfai2014}, it is assumed that the state of a node will be lost if it does not interact with others or itself. This difference implies that state persistent links are conceptually different from the self-interactions or self-loops as mentioned in \cite{Posfai2014}. For state persistent links, their values are always 1. However, for self-interactions, their values can be different from 1 and measured from data.

\subsection{Calculating the controllable system size $N_b$}
\begin{figure}
	\centering
	\includegraphics[width=0.8\textwidth]{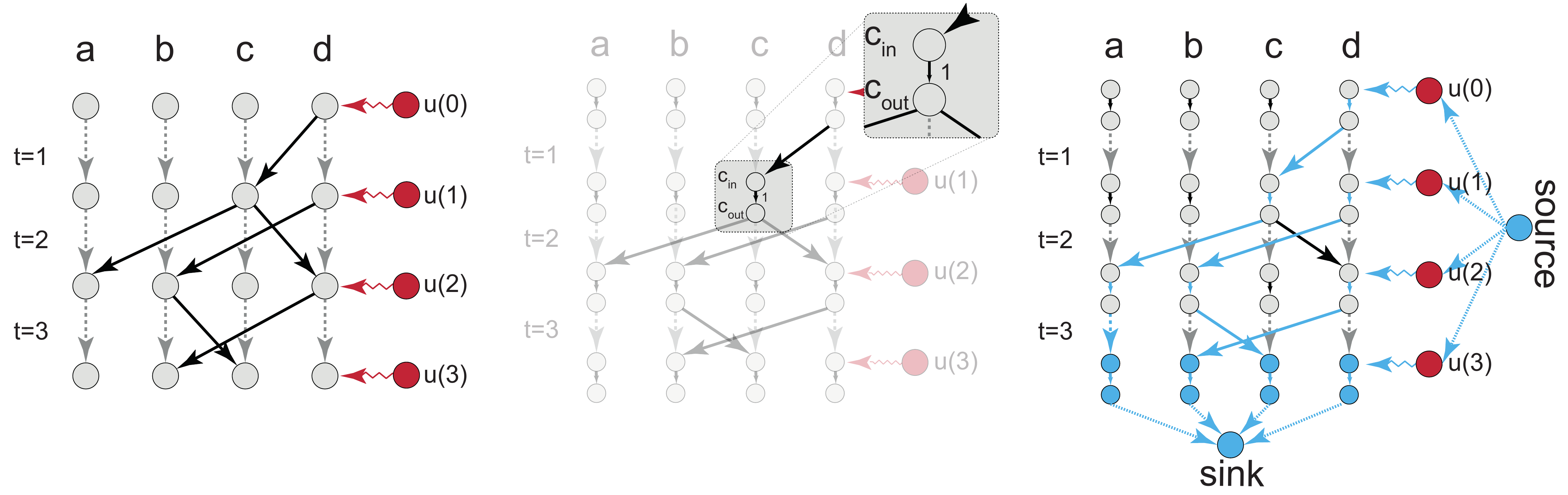}

	\caption{Illustration of the auxiliary time-unfolded network to identify the maximum number of independent time-respecting paths. This illustration shows the case where an input signal is attached to only one driver node, however the same construction applies to cases with multiple driver nodes.}
	\label{aux}
\end{figure}

We calculate the controllable system size $N_b$ by identifying the maximum number of independent paths in a time-unfolded network.
The procedure works by constructing an auxiliary network $H$.
First, we replace each node $v$ except for driver nodes with $v_{out}$ and $v_{in}$. (see Figure.~\ref{aux}(a)) where $v_{in}$ collects all links pointing to $v$ while $v_{out}$ collects all links originating from $v$.
We further include an additional link from each $v_{in}$ to the corresponding $v_{out}$. This node-splitting procedure reflects the constraint that two paths can not pass through the same node $v$ if we set the weight of this additional link to $1$.
Moreover, we add one \emph{source node} which is connected via directed links to all input signals at all time steps.
Finally, we add one \emph{sink node} along with directed links connecting all temporal copies at time $T$ to this sink node.
The result is the auxiliary network $H$ presented in Figure.~\ref{aux} (c).
Based on this construction, the task of finding a maximum set of independent time-respecting paths corresponds to identifying a maximum flow from source to sink in the auxiliary network where all link capacities are set to one\cite{Liu2011}.
These link capacities of one capture the constraint that only one path is allowed to pass through one node at a given time.
With this network $H$, the size of the controllable subsystem $N_b$ at time $T$ corresponds to the maximum flow from source to sink, which can be easily solved in polynomial time~\cite{Goldberg1988}.

\bibliographystyle{unsrt}

\bibliography{temporalcontrol}

\end{document}